\documentclass{aa}

\usepackage{graphicx}
\usepackage{txfonts}
\usepackage{hyperref}
\usepackage{amsmath}
\usepackage[noabbrev]{cleveref}
\begin{document} 

   \title{On-sky verification of Fast and Furious focal-plane wavefront sensing: Moving forward toward controlling the island effect at Subaru/SCExAO}
   
   \author{S.P. Bos\inst{1},  S. Vievard\inst{2,3,4}, M.J. Wilby\inst{1}, F. Snik\inst{1}, J. Lozi\inst{2}, O. Guyon\inst{2,4,5,6}, B. R. M. Norris\inst{7,8,9}, N. Jovanovic\inst{10}, F. Martinache\inst{11}, J.-F. Sauvage\inst{12,13}, and C.U. Keller\inst{1}}
   \institute{Leiden Observatory, Leiden University, P.O. Box 9513, 2300 RA Leiden, The Netherlands
                \and    
                National Astronomical Observatory of Japan, Subaru Telescope, National Institute of Natural Sciences, Hilo, HI 96720, USA
                \and
                Observatoire de Paris - LESIA, 5 Place Jules Janssen, 92190 Meudon, France
                \and
                Astrobiology Center, National Institutes of Natural Sciences, 2-21-1 Osawa, Mitaka, Tokyo, Japan
                \and
                Steward Observatory, University of Arizona, 933 N. Cherry Ave, Tucson, AZ 85721, USA
                \and
                College of Optical Sciences, University of Arizona, 1630 E. University Blvd., Tucson, AZ 85721, USA
                \and 
                Sydney Institute for Astronomy, School of Physics, Physics Road, University of Sydney, NSW 2006, Australia
                \and
                Sydney Astrophotonic Instrumentation Laboratories, Physics Road, University of Sydney, NSW 2006, Australia 
                \and
                Australian Astronomical Observatory, School of Physics, University of Sydney, NSW 2006, Australia 
                \and
                Department of Astronomy, California Institute of Technology, 1200 E. California Blvd., Pasadena, CA 91125, USA   
                \and
                Observatoire de la Cote d'Azur, Boulevard de l'Observatoire, Nice, 06304, France
                \and
                Aix Marseille Univ, CNRS, LAM, Laboratoire d'Astrophysique de Marseille, Marseille, France
               \and 
               ONERA, 29 Avenue de la Division Leclerc, 92320 Ch\^{a}tillon, France\\     
               \email{stevenbos@strw.leidenuniv.nl}
             }

   \date{Received March 9, 2020; accepted May 18, 2020}

 
  \abstract
   {High-contrast imaging (HCI) observations of exoplanets can be limited by the island effect (IE). 
    The IE occurs when the main wavefront sensor (WFS) cannot measure sharp phase discontinuities across the telescope's secondary mirror support structures (also known as spiders). 
    On the current generation of telescopes, the IE becomes a severe problem when the ground wind speed is below a few meters per second.
    During these conditions, the air that is in close contact with the spiders cools down and is not blown away. 
    This can create a sharp optical path length difference (OPD) between light passing on opposite sides of the spiders. 
    Such an IE aberration is not measured by the WFS and is therefore left uncorrected.
    This is referred to as the low-wind effect (LWE). 
    The LWE severely distorts the point spread function (PSF), significantly lowering the Strehl ratio and degrading the contrast.
    }
   {In this article, we aim to show that the focal-plane wavefront sensing (FPWFS) algorithm, Fast and Furious (F\&F), can be used to measure and correct the IE/LWE. 
   The F\&F algorithm is a sequential phase diversity algorithm and a software-only solution to FPWFS that only requires access to images of non-coronagraphic PSFs and control of the deformable mirror. 
   }
   {We deployed the algorithm on the SCExAO HCI instrument at the Subaru Telescope using the internal near-infrared camera in H-band. 
    We tested with the internal source to verify that F\&F can correct a wide variety of LWE phase screens. 
    Subsequently, F\&F was deployed on-sky to test its performance with the full end-to-end system and atmospheric turbulence.
   The performance of the algorithm was evaluated by two metrics based on the PSF quality: 1) the Strehl ratio approximation ($SRA$), and 2) variance of the normalized first Airy ring ($VAR$).
   The $VAR$ measures the distortion of the first Airy ring, and {is used to quantify PSF improvements that do not or barely affect the PSF core} (e.g., during challenging atmospheric conditions).  
   }
   {The internal source results show that F\&F can correct a wide range of LWE phase screens. 
   Random LWE phase screens with a peak-to-valley wavefront error between 0.4 $\mu$m and 2 $\mu$m were all corrected to a $SRA$ $>$90\% and an $VAR\lessapprox0.05$. 
   Furthermore, the on-sky results show that F\&F is able to improve the PSF quality during very challenging atmospheric conditions {(1.3-1.4'' seeing at 500 nm)}. 
   Closed-loop tests show that F\&F is able to improve the $VAR$ from 0.27 to 0.03 and therefore significantly improve the symmetry of the PSF.
   Simultaneous observations of the PSF in the {optical} ($\lambda = $ 750 nm, $\Delta \lambda =$ 50 nm) show that during these tests              {we were correcting aberrations common to the optical and NIR paths within SCExAO.} 
   {We could not conclusively determine if we were correcting the LWE and / or (quasi-)static aberrations upstream of SCExAO.}
   }
   {The F\&F algorithm is a promising focal-plane wavefront sensing {technique} that has now been successfully tested on-sky. 
   Going forward, the algorithm is suitable for incorporation into observing modes, which will enable {PSFs of higher quality and stability} during science observations.}


\titlerunning{On-sky verification of ``Fast \& Furious'' focal-plane wavefront sensing}
\authorrunning{S. P. Bos et al.}
\maketitle

\section{Introduction}\label{sec:introduction}
Current high-contrast imaging (HCI) instruments, such as SCExAO \citep{jovanovic2015subaru}, MagAO-X (\citealt{males2018magao}; \citealt{close2018optical}), SPHERE \citep{beuzit2019sphere}, and GPI \citep{macintosh2014first}, are now routinely exploring circumstellar environments at high contrast ($\sim$$10^{-6}$) and small angular separation ($\sim$$200$ mas) {in the near-infrared or the optical
} \citep{vigan2015high}. 
These instruments detect and characterize exoplanets by means of direct imaging, integral field spectroscopy, or polarimetry (\citealt{macintosh2015discovery}; \citealt{keppler2018discovery}). 
Such observations help us to understand the orbital dynamics of planetary systems \citep{wang2018dynamical}, the composition of the exoplanet's atmosphere \citep{hoeijmakers2018medium}, and find cloud structures \citep{stam2004using}.
To reach these extreme contrasts and angular separations, these instruments use extreme adaptive optics to correct for turbulence in the Earth's atmosphere, coronagraphy to remove unwanted star light, and advanced post-processing techniques to enhance the contrast, for example, angular differential imaging \citep{marois2006angular}, reference star differential imaging \citep{ruane2019reference}, spectral differential imaging \citep{sparks2002imaging}, and polarimetric differential imaging (\citealt{langlois2014high} ; \citealt{van2017combining}). \\
\indent One of the limitations of the current generation of HCI instruments are aberrations that are non-common and chromatic between the main wavefront sensor arm and the science focal-plane. 
These non-common path aberrations (NCPA) vary on minute to hour timescales during observations, due to a changing gravity vector, humidity, and temperature (\citealt{martinez2012speckle}; \citealt{martinez2013speckle}), and are therefore difficult to remove in post-processing. 
Ideally, these aberrations are detected by wavefront sensors close to, or in the science focal plane and subsequently corrected by the deformable mirror (DM).  
Many variants of such wavefront sensors have been developed, and some of these have been successfully demonstrated on-sky (\citealt{martinache2014sky}; \citealt{singh2015sky}; \citealt{martinache2016closed}; \citealt{bottom2017speckle}; \citealt{wilby2017coronagraphic}; \citealt{bos2019focal}; \citealt{galicher2019minimization}; \citealt{vigan2019calibration}).\\

\begin{figure}
\centering
\includegraphics[width=\hsize]{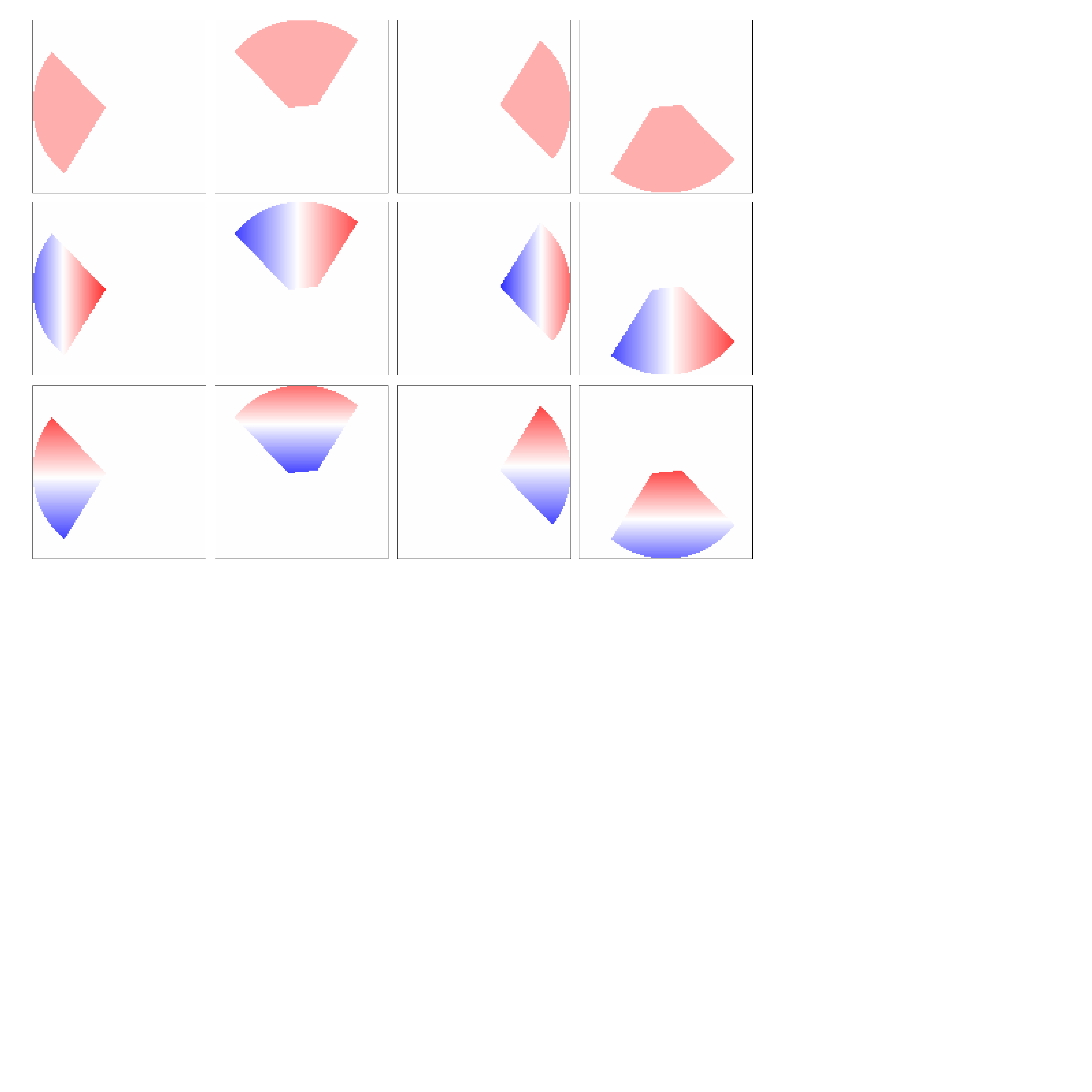}
\caption{Piston-tip-tilt mode basis for SCExAO instrument at the Subaru Telescope. 
             The pupil of SCExAO is fragmented into four segments due to the spiders, see \autoref{fig:subaru_pupil}.
             For every individual segment, we define a piston, tip, and tilt mode. 
             }
\label{fig:PTT_modes}
\end{figure}

\indent Another limitation is the island effect (IE), which occurs when the telescope pupil is strongly fragmented by support structures for the secondary mirror. 
We refer to these fragments as segments in the rest of the paper. 
When these structures become too wide, conventional pupil-plane wavefront sensors (WFSs) such as the Shack-Hartmann and Pyramid poorly sense sharp discontinuities in phase aberrations across these gaps. 
This is because these WFSs typically measure the gradient of the wavefront in two orthogonal directions, and discontinuities can be difficult to integrate over to retrieve the wavefront itself.
It is expected that the upcoming class of Giant Segmented Mirror Telescopes (GSMTs) will increasingly suffer from the IE, as the support structures will become even wider and more numerous.\\ 

\indent For the current generation of HCI instruments, the IE {mainly manifests itself as the so-called low-wind effect (LWE).} 
The LWE occurs when the ground windspeed is very low (under a few m/s), which would typically be considered {to be amongst the best} observing conditions. 
It has now been well understood to be a form of dome seeing and is caused by thermal problems at the spiders supporting the secondary mirror (\citealt{sauvage2015low}; \citealt{sauvage2016tackling}; \citealt{milli2018low}). 
During these events, radiative cooling of the spiders lowers their temperature below that of the ambient air. 
The air on one side of the spider that is in close contact, and which is not blown away due to the low wind speeds, also cools down and changes its refractive index.
This introduces a sharp optical path length difference (OPD) between light passing on opposite sides of a spider, which is subsequently not measured by the traditional wavefront sensor. 
The aberrations generated by the LWE were measured to have a peak-to-valley (P-V) wavefront error (WFE) of up to hundreds of nanometers \citep{sauvage2015low}, and can be considered to be a combination of piston-tip-tilt (PTT) phase modes across each segment. 
We invite the reader to see \autoref{fig:PTT_modes} for an example of such modes in the context of the Subaru Telescope pupil. 
Typical consequences of the LWE are a strong distortion of the point spread function (PSF), the first Airy ring broken up into multiple side lobes, and an accompanying strong reduction in Strehl ratio (typically tens of percent).
This results in a reduced relative signal from circumstellar objects and degraded raw contrasts, and thus an overall worse performance of the HCI system.
Furthermore, these effects are generally quasi-static and thus become difficult to calibrate in post-processing. 
The LWE has been reported at the {VLT and Subaru telescopes to affect 3\% to 20\% of the observations, while Gemini South is at $<3$\%} \citep{milli2018low}. \\

\indent Thus far, multiple solutions have been investigated that either prevent the LWE from occurring, or measure it with an additional wavefront sensor and correct it with the DM. 
At the VLT, the spiders were recoated with a {material} that has a low thermal emissivity in the infrared. 
This brought the occurrence rate down from 20\% to a more manageable 3\% \citep{milli2018low}. 
But it is still reported when the ground wind speed is below 1 m/s, making additional solutions that drive this down even further  desirable. 
In the context of future instruments of GSMTs, there have also been investigations \citep{hutterer2018advanced} toward changing the wavefront reconstruction of the Pyramid WFS to make it sensitive to the IE and therefore the LWE.
Several focal-plane wavefront sensors have also been investigated to specifically target the LWE. 
For example, the Asymmetric Pupil Fourier Wavefront Sensor (APF-WFS; \cite{martinache2013asymmetric}) was demonstrated on-sky at Subaru/SCExAO to be able to correct the LWE \citep{n2018calibration}.  
At Subaru/SCExAO, a host of new focal-plane wavefront sensing methods are being tested with the internal source and on-sky in the context of the IE and LWE \citep{vievard2019overview}. \\

\indent In this paper, we present the results of deploying one of these methods, the Fast and Furious algorithm (F\&F; \citealt{keller2012extremely}; \citealt{korkiakoski2012experimental}; \citealt{korkiakoski2014fast}), to the SCExAO instrument.
This algorithm is a software-only solution to focal-plane wavefront sensing and therefore easy to implement on HCI instruments. 
It will be more extensively discussed in \autoref{sec:theory}. 
In previous work, F\&F was already explored as a way to measure the LWE in the context of the SPHERE instrument (\citealt{wilby2016fast}; \citealt{wilby2018laboratory}).
Specifically, the goal was to show that the algorithm would still perform well in the low signal-to-noise environment of the differential tip-tilt sensor \citep{baudoz2010differential} within SPHERE. 
It showed satisfactory performance both in simulation \citep{wilby2016fast} and at the MITHIC bench \citep{vigan2016characterisation} in a laboratory environment \citep{wilby2018laboratory}.
Here, we study the performance of the algorithm on the SCExAO instrument using the internal source, and report on the first on-sky tests in \autoref{sec:demonstration}.  
We discuss the results and conclude in \autoref{sec:conclusions}.

\section{Fast and Furious algorithm}\label{sec:theory} 
\begin{figure}
\centering
\includegraphics[width=\hsize]{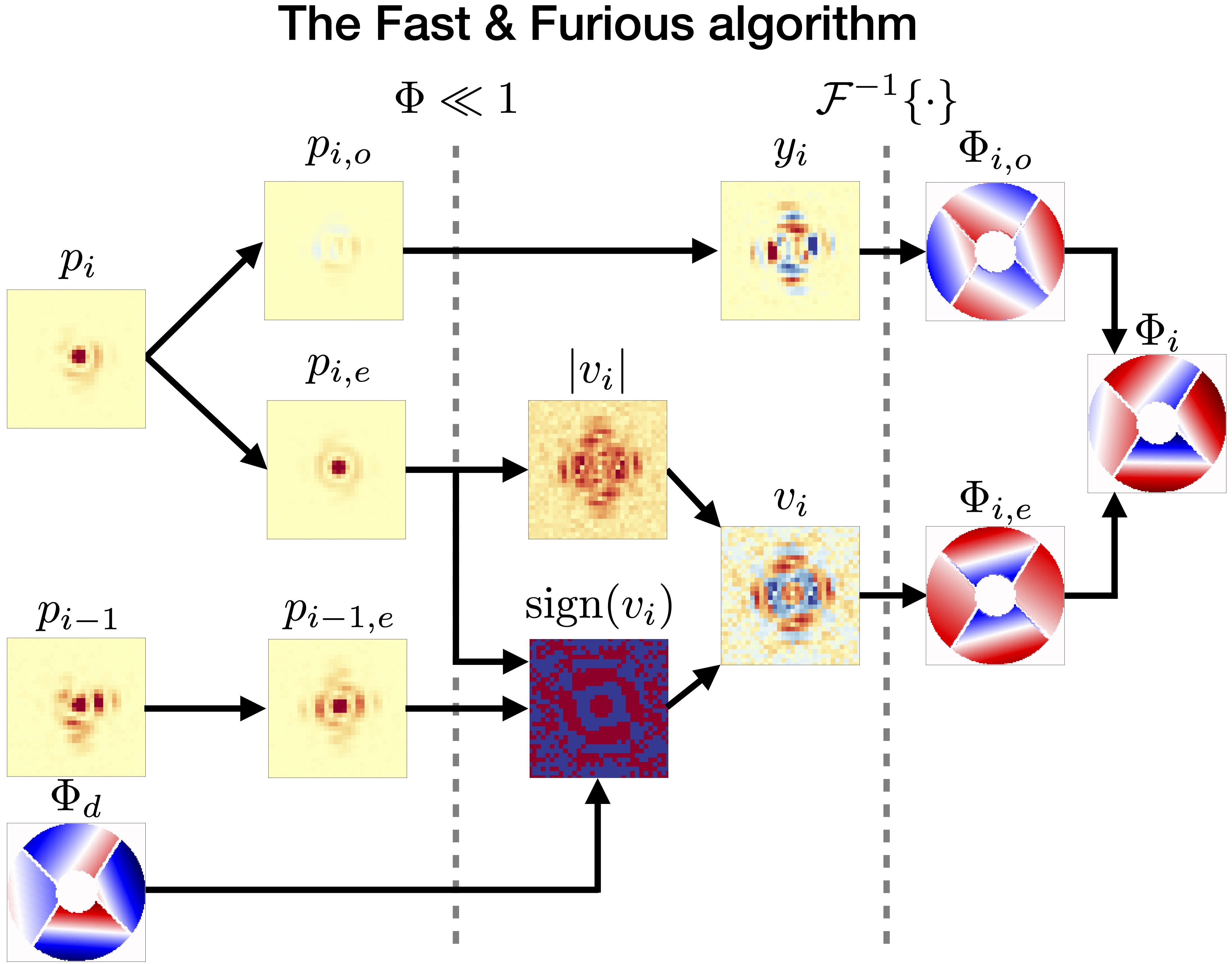}
\caption{Explanation of one iteration of the Fast and Furious algorithm. At iteration $i$ an image $p_i$ is split into its even $p_{i,e}$ and odd $p_{i,o}$ components. The odd component can directly solve for the odd focal-plane electric field $y_i$ (\autoref{eq:odd_electric_field}). Similarly, the even component is used to solve for the absolute value of the even focal-plane electric field $|v_i|$ (\autoref{eq:abs_value_even_electric_field}). To solve for the sign of $v_i$, the previous image $p_{i-1}$, that has a diversity phase $\Phi_d$, is introduced to break the degeneracy (\autoref{eq:sign_even_electric_field}). The estimates of $y_i$ and $v_i$ together give an estimate of the pupil-plane phase $\Phi_i$ (\autoref{eq:phase_estimate}).}
\label{fig:FF_explanation}
\end{figure}

The Fast and Furious (F\&F; \citealt{keller2012extremely}; \citealt{korkiakoski2014fast}) algorithm is an extension of the sequential phase diversity technique originally introduced by \cite{gonsalves2002adaptive}. 
In conventional phase diversity techniques (\citealt{gonsalves1982phase}; \citealt{paxman1992joint}), the degeneracy in estimating even phase modes is solved by recording two images, one in focus and another strongly out of focus. 
This forces the user to either split the light into two imaging channels or alternately record in- and out-of-focus images. 
A sequential phase diversity algorithm uses sequential in-focus images and relies on a closed-loop system that continuously provides phase corrections that improve the wavefront and serve as diversity to solve for the even phase aberrations. 
Therefore, such an algorithm will never be able to give a single shot phase estimate and must always be operated in closed loop. \\

\indent The F\&F algorithm refers to an extension of this sequential phase diversity technique and greatly improves the dynamic range and stability \citep{keller2012extremely}. 
Focal-plane images acquired by the algorithm are split into the even and odd components. 
Using simple algebra, the odd component directly solves for the odd focal-plane electric field.  
The even component can only solve for the absolute value of the even focal-plane electric field. 
To acquire the sign of the even electric field, F\&F uses the image and change in the phase introduced by the DM of the previous iteration to break the degeneracy. 
Together, these operations give an estimate of focal-plane electric field, and, by an inverse Fourier transformation, an estimate of the pupil-plane phase. 
As one F\&F iteration only relies on simple algebra and a single Fourier transformation, the algorithm is computationally very efficient, and can in principle run at high frame rates.\\ 

\indent An extensive discussion on the algorithm and its performance is presented in \cite{keller2012extremely} and \cite{korkiakoski2014fast}. 
Here, we give an overview of the key F\&F equations that lead to a phase estimate. 
A graphical overview of the algorithm is shown in \autoref{fig:FF_explanation}.
{For these equations, we notably assume; (i) real and symmetric pupil amplitude (which is a reasonable assumption for most telescope and instrument pupils); (ii) monochromatic light (performance of the algorithm decreases when the bandwidth increases); (iii) phase-only aberrations (an extension of F\&F deals with amplitude aberrations \citep{korkiakoski2014fast}); {and} (iv) phase aberrations can be approximated to be small ($\Phi \ll 1$ radian).}
The point-spread-function (PSF) of an optical system is given by: 
\begin{equation}
p = |\mathcal{F}\{\ A e^{i \Phi} \}|^2.
\end{equation}
Here, $p$ is the PSF, $A$ and $\Phi$ the pupil-plane amplitude and phase, and $\mathcal{F}\{\cdot\}$ the Fourier transformation operator.
For F\&F, the assumption is that $A$ is real and symmetric. 
We adopt the same notation as in \cite{wilby2018laboratory}, which means that pupil-plane quantities are denoted by upper case variables and focal-plane quantities by lower case variables.
Assuming that $\Phi \ll 1$, we can expand the PSF to second order, which results in: 
\begin{equation}\label{eq:PSF_approx}
p \approx S a^2 + 2 a (ia * \phi_o) + (ia * \phi_o)^2 + (a * \phi_e)^2.
\end{equation}
With the electric field of the unaberrated PSF given by $a = \mathcal{F}\{A\}$, the Fourier transforms of the even and odd pupil-plane phases ($\Phi = \Phi_o + \Phi_e$) are given by $\phi_o = \mathcal{F}\{\Phi_o\}$ and $\phi_e = \mathcal{F}\{\Phi_e\}$. 
The normalization factor $S = 1 - \sigma_{\phi}^2$ can be understood as the first order Mar\'{e}chal approximation of the Strehl ratio \citep{roberts2004really}, with $\sigma_{\phi}^2$ the wavefront variance. 
This approximation becomes highly accurate when the aberrations are small. 
The convolution operator is denoted by $*$.
It is more convenient to express \autoref{eq:PSF_approx} in terms of the odd and even focal-plane electric fields, which are given by: 
\begin{align}
y &= i \mathcal{F}\{ A \Phi_o \} = ia * \phi_o, \\
v &= \mathcal{F}\{ A \Phi_e \} = a * \phi_e.
\end{align}
Splitting the PSF (\autoref{eq:PSF_approx}) in its odd and even components ($p = p_o + p_e$), and solving for $y$ and $v$ results in: 
\begin{align}
y   &= a p_o / (2 a^2 + \epsilon), \label{eq:odd_electric_field} \\
|v| &= \sqrt{|p_e - (Sa^2 + y^2)|} \label{eq:abs_value_even_electric_field}.
\end{align}
Here, $\epsilon$ is a regularization parameter for the pixels where $a$ goes to zero that would otherwise amplify the noise. 
This solution only solves for $|v|$, which is a well-known sign ambiguity (\citealt{gonsalves1982phase}; \citealt{paxman1992joint}).
The sign of $v$ is solved by introducing an additional image that has a known phase diversity $\Phi_d$. 
This additional image is for F\&F the image of the previous iteration; 
because it has a phase diversity with respect to the current iteration, given by the change in DM command (assuming that $\Phi$ remains constant).
The PSFs of these two images can be approximated by:
\begin{align}
p_i &\approx S a^2 + 2 a y + y^2 + v^2, \\
p_{i-1} &\approx S a^2 + 2 a (y + y_d) + (y + y_d)^2 + (v + v_d)^2,
\end{align}
with $y_d = i \mathcal{F}\{ A \Phi_{d,o} \}$ and $ v_d = \mathcal{F}\{ A \Phi_{d,e} \}$ the odd and even focal-plane electric fields of the diversity. 
It is most robust to estimate only the sign of $v$ (instead of the complete $v$) by:
\begin{equation} \label{eq:sign_even_electric_field}
\text{sign}(v) = \text{sign} \left( \frac{p_{i-1,e}  - p_{i,e} - (v_d^2 + y_d^2  + 2 y y_d)}{2 v_d} \right).
\end{equation}
For the first iteration of F\&F, when there is no diversity image available, the most optimal guess is $\text{sign}(v)=a$. 
Although this guess might be wrong, it will provide sufficient diversity to make the following estimates of the even wavefront accurate. 
The estimate of the odd part of the wavefront is unaffected by any sign error, and therefore will be improved from the first iteration.
The final pupil-plane phase estimate for this iteration is given by: 
\begin{equation}\label{eq:phase_estimate}
A \Phi = \mathcal{F} ^{-1} \{ \text{sign}(v) |v| - i y \}
.\end{equation}
This phase estimate can be subsequently projected onto a mode basis of choice to target specific aberrations. For example, the piston-tip-tilt (PTT) mode basis shown in \autoref{fig:PTT_modes} is designed specifically for the LWE, {and/or} the lowest Zernike modes for NCPA caused by optical misalignments \citep{wilby2018laboratory}.
\section{Demonstration at Subaru/SCExAO}\label{sec:demonstration}
\subsection{SCExAO and algorithm implementation}\label{sec:implementation}
\begin{figure}
\centering
\includegraphics[width=\hsize]{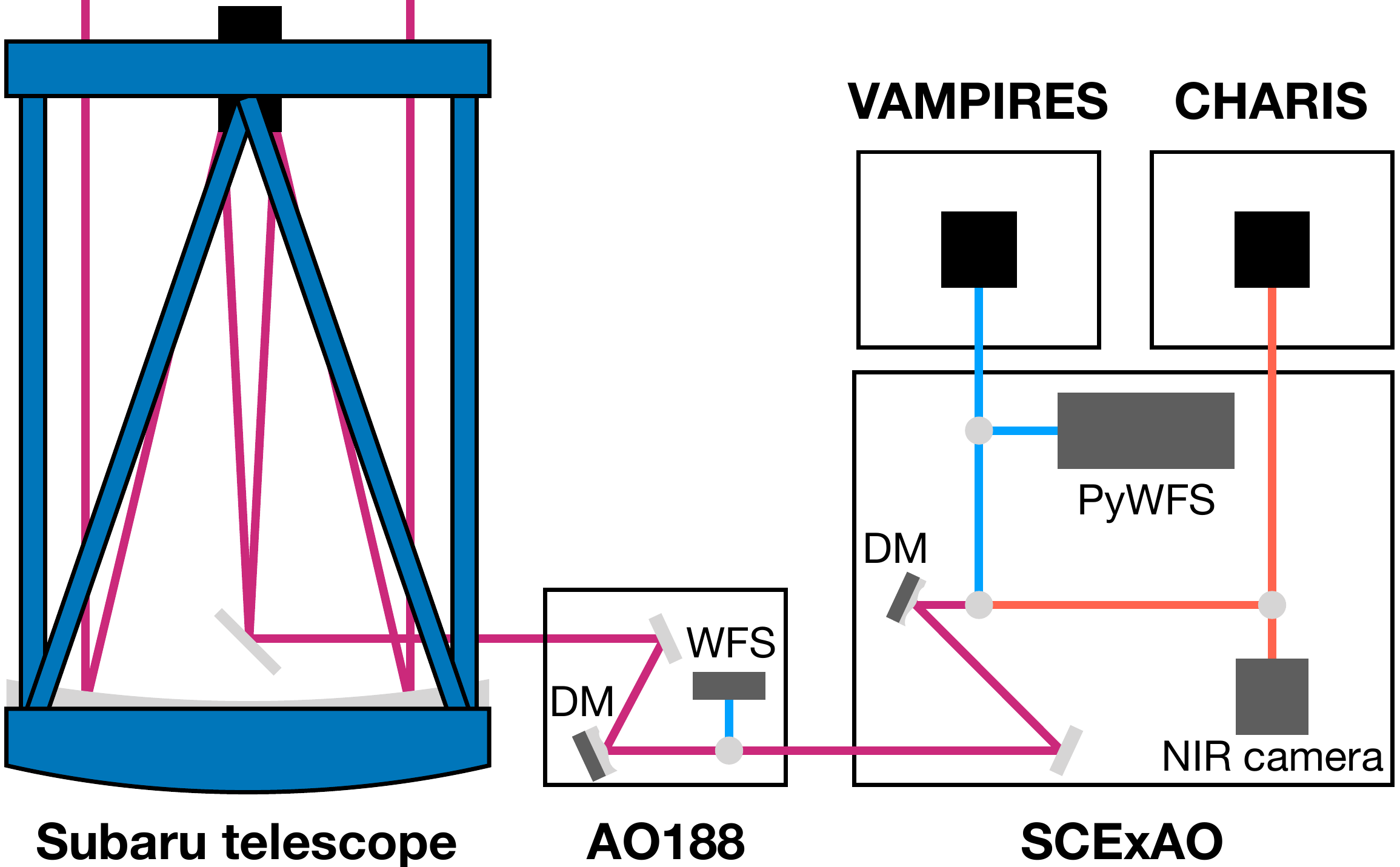}
\caption{Schematic of the complete system layout. 
             Acronyms in the figure are: deformable mirror (DM), wavefront sensor (WFS),  pyramid wavefront sensor (PyWFS), near infrared (NIR).}
\label{fig:system_layout}
\end{figure}
\begin{figure}
\centering
\includegraphics[width=\hsize]{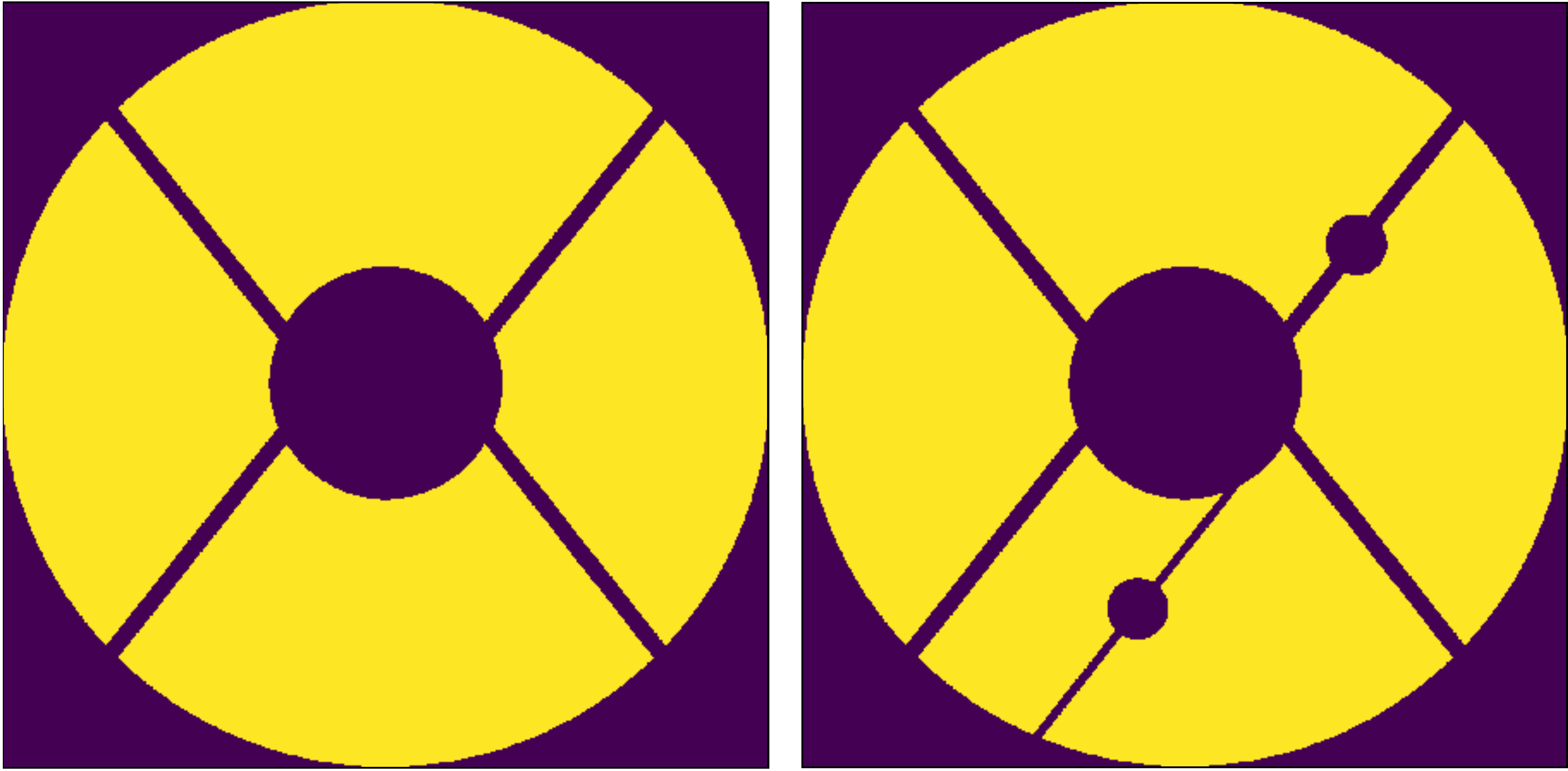}
\caption{Pupil of Subaru pupil (left), and the SCExAO instrument (right). The SCExAO pupil has additional structure to block unresponsive actuators in the deformable mirror. {The spiders are 23 cm wide and up to $\sim$1 m high \citep{milli2018low}.}}
\label{fig:subaru_pupil}
\end{figure}
We deployed F\&F to the Subaru Coronagraphic Extreme Adaptive Optics (SCExAO) instrument \citep{jovanovic2015subaru}, which is located on the Nasmyth platform of the Subaru Telescope downstream of the AO188 system \citep{minowa2010performance}. 
We invite the reader to see \autoref{fig:system_layout} for a schematic of the telescope, AO188, and SCExAO. 
The main wavefront sensor in the instrument is a pyramid wavefront sensor (PYWFS; \citealt{lozi2019visible}) in the 600-950 nm wavelength range. 
The real-time control is handled by the Compute And Control for Adaptive Optics (CACAO) software package \citep{guyon2018compute} that sends the wavefront corrections to the 2000-actuator deformable mirror (DM). 
The active pupil on the DM has a diameter of 45 actuators, which gives SCExAO a control radius of 22.5 $\lambda/D$. 
The CACAO software allows for additional wavefront corrections to be sent by other wavefront sensors, by treating their corrections on separate DM channels.
It updates the PYWFS reference offset to make sure that the AO loop does not cancel commands of the other wavefront sensors.
The current science modules fed by SCExAO are VAMPIRES \citep{norris2015vampires} in the optical, and CHARIS (\citealt{peters2013optical}; \citealt{groff2014construction}) in the near-infrared, but more are foreseen (\citealt{lozi2018scexao}; \citealt{lozi2019new}; \citealt{guyon2019scexao}). \\ 
\indent The F\&F algorithm is implemented using Python and the HCIPy package \citep{por2018hcipy}.
Python has a simple interface with the instrument and allows for rapid testing. 
We tested F\&F using the internal NIR C-RED 2 camera that has a $640\times512$ pixel InGaAs sensor cooled to $-40$ $^{\circ}$ C \citep{feautrier2017c}.
The images were cropped to $64\times64$ pixels, dark-subtracted, flat-fielded and subsequently aligned with a reference PSF from a numerical model. 
Alignment with the reference PSF {improved} the stability of F\&F (or any other FPWFS algorithm), but therefore {tip and tilt were no longer measured}.  
The number of images stacked for one F\&F iteration {was} generally between 1 and 100. 
The algorithm was tested using a narrowband filter ($\Delta \lambda = 25$ nm) at 1550 nm and the H-band filter.
We note that the quantum efficiency of the detector in the C-RED 2 camera rapidly {decreases when the wavelength is} above $\sim$1630 nm, and therefore the tests using the H-band filter only {used} approximately half of the wavelength range.
As explained in \autoref{sec:theory}, F\&F assumes a real and symmetric pupil amplitude.
In \autoref{fig:subaru_pupil}, we show in the left subfigure the nominal Subaru pupil and on the right subfigure the SCExAO pupil.
SCExAO defines its pupil internally, because there are unresponsive actuators in the DM that need to be blocked, which is shown in the figure.
Normally, we would {have used} the right subfigure to be the pupil amplitude for F\&F, and accept that the relatively small asymmetry would introduce a bias in the wavefront estimate. 
But, during the tests presented in this paper, this internal mask in SCExAO was damaged (the structure blocking the dead actuator in the lower segment was broken off), and thus we assumed the nominal Subaru pupil for $A$.
To accurately calculate $a = \mathcal{F}\{ A\}$ on the detector, we {had} to take into account the plate scale and the rotation of the pupil with {respect} to the detector. 
We determined these parameters by fitting a simple model of the PSF to data from the instrument, using the pupil in \autoref{fig:subaru_pupil} and the rotation and plate scale as free parameters. 
This resulted in a plate scale of $15.45$ mas/pixel and a counterclockwise rotation of $9.6^{\circ}$. 
As discussed in \autoref{sec:theory}, the phase estimate by F\&F as shown in \autoref{eq:phase_estimate} can be projected on a mode basis. 
This can have multiple advantages: first, if the goal is to just control a certain mode basis (e.g., the PTT modes or low-order Zernike modes for NCPA); second, by filtering out the (noisier) higher spatial frequency modes, the noise in the phase estimate is reduced; and third, removing any systematics due to inaccurate pupil symmetry assumptions.
As the goal of this paper {was} to measure the LWE using a camera downstream of the PYWFS, we projected the phase estimates of F\&F on a mode basis that consisted of the PTT modes shown in \autoref{fig:PTT_modes} and/or the lowest 50 Zernike modes (starting at defocus) for NCPA estimation .
We did not estimate tip and tilt, because all images are aligned with a reference PSF. 
The combined PTT and Zernike mode basis {was not orthogonalized}, and therefore there {could have been} some cross-talk. 
{However, as we operated in closed loop, we initially expected these effects to be minimal, and in the end did not notice any significant effects.}
The algorithm {used} its own phase estimates (after the decomposition; multiplied by the loop gain) for the phase diversity. 
Estimates of PYWFS would not be useful as they will not see the same aberration due to NCPA, chromatic effects, and the null-space of the PYWFS. 
The DM command $\theta_{\text{DM},i}$ at iteration $i$ sent to CACAO by F\&F for wavefront control {was} calculated by:
\begin{equation}
\theta_{\text{DM},i} = c_{lf}\theta_{\text{DM},i-1} - \frac{g}{2} \Phi_i, 
\end{equation}
with $g$ the loop gain (mostly set between $0.1$ and $0.3$), and $c_{lf}$ the leakage factor (generally between 0.99 and 0.999).
The factor $\frac{1}{2}$ {was} to account for the reflection of the DM, and $\Phi_i$ the phase estimate by F\&F at iteration $i$.
{We {computed} the DM commands as actuator displacements in micrometers, which {were} converted to voltages internally by CACAO.}
The loop speed {during the tests presented in this work was} generally between 4 and 25 frames per second (FPS), {and depends} on the image size, the number of images stacked {($N_{\text{img avg}}$),} and the size of the mode basis on which the phase estimate is decomposed. 
Currently, {the main limitation is $N_{\text{img avg}}$, because each of the images needs to be aligned, which is the most time-consuming process.} 
{The image alignment code uses the Python library Scipy \citep{jones2014scipy}.}
It is expected that if the algorithm (including the image alignment routines) were {completely} written in C (used by CACAO), 300 - 400 FPS would be relatively easily to achieve if that is desirable. 
\subsection{Quantifying PSF quality}
We {quantified} the quality of the PSF by the Strehl ratio approximation. 
The Strehl ratio approximation ($SRA$) is estimated by comparing the data $p$ with a numerical PSF $|a|^2$ (that has been oversampled by a factor of 16) by using a modified encircled energy metric: 
\begin{equation}\label{eq:strehl_measurement}
SRA = \frac{p(r < 1.22 \ \lambda / D)}{p(r < 11.5 \ \lambda / D)} \cdot \frac{|a|^2(r < 11.5 \ \lambda / D)}{|a|^2(r < 1.22 \ \lambda / D)}
.\end{equation}
{The $SRA$ is calculated at $\lambda = 1550$ nm.}
We note that it is very difficult to make an accurate Strehl measurement \citep{roberts2004really}, for example, {in} our metric aberrations that impact the PSF beyond 11.5 $\lambda / D$ are not taken into account.
Furthermore, as all images are aligned with a numerical reference PSF, the brightest peak of a severely distorted image will be aligned with the PSF core. 
This means that images with a low Strehl ratio ($\sim$0-50\%) {are} reported with a much higher $SRA$.  
We chose this metric over residual wavefront measurements, because there was not an independent WFS available that is sufficiently common-path with the C-RED 2 camera during either internal source or on-sky tests. 
Furthermore, at high Strehl ratios it is still a good indication of residual wavefront variance. \\
\indent Some of the on-sky results were taken during challenging atmospheric conditions, for example during the tests on December 12, 2019, we recorded a 1-1.1" seeing in H-band{, corresponding to 1.3-1.4" seeing at 500 nm\footnote{The seeing scales with $\lambda^{-1/5}$ \citep{hardy1998adaptive}.}} . 
This meant that when the F\&F loop was closed, the PSF would qualitatively improve (it became more symmetric), but the improvement was not reflected in an increased $SRA$. 
{Therefore, we defined} a metric that measures the quality of the first Airy ring, because the low-order nature of LWE aberrations results in strong distortions of the first Airy ring and it is easy to measure. 
The Variance of the normalized first Airy ring ($VAR$) is defined as:
\begin{multline}\label{eq:VAR}
VAR = \text{Var} \biggl( \frac{p(1.52 \lambda / D < r < 2.14 \lambda / D)}{\langle p(1.52 \lambda / D < r < 2.14 \lambda / D) \rangle} \cdot \\
\frac{\langle |a|^2(1.52 \lambda / D < r < 2.14 \lambda / D) \rangle}{|a|^2(1.52 \lambda / D < r < 2.14 \lambda / D)}  \biggr) 
\end{multline}  
We only {select} the peak of the Airy ring (i.e., $1.52 \lambda / D$$<$$r$$<$$2.14 \lambda / D$), as that is where the effects are the strongest. 
Furthermore, the Airy ring is normalized twice, first by its mean in order for us to measure relative disturbances.
And subsequently, by the normalized Airy ring of a numerically calculated PSF. 
This is necessary because there are natural variations in brightness across the Airy ring due to the diffraction structures of the spiders that we want to divide out. 
An undistorted PSF will therefore have $VAR$$=$$0$, while distorted PSFs will have $VAR$$>$$0$.
Based on the experiments with the internal source, {a VAR of 0.03-0.05 can be considered as} good. 
We note that the $VAR$ is insensitive to aberrations that are azimuthally symmetric, for example, defocus and spherical aberration, as these are be removed by the first normalization step.
\subsection{Internal source demonstration}\label{sec:internal_source} 
\begin{figure*}
\centering
   \includegraphics[width=17cm]{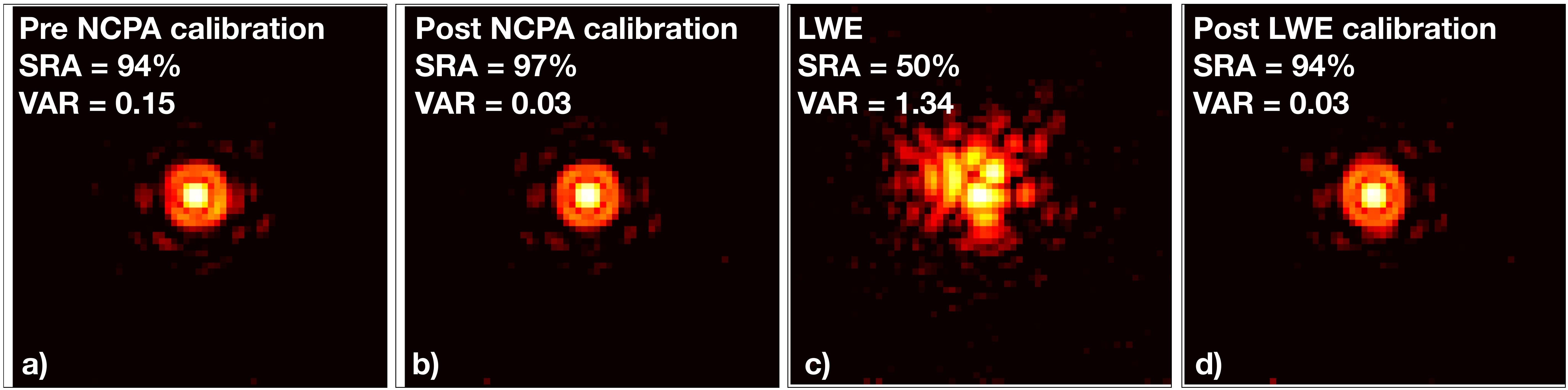}
     \caption{Images during tests with the internal source. PSFs are normalized to their maximum value, {and are plotted in logarithmic scale}. PSF before (a) and after (b) NCPA calibration. Introduction of the LWE phase screen (c) and the PSF after correction (d).}
     \label{fig:internal_source_images}
\end{figure*} 
\begin{figure*}
\centering
   \includegraphics[width=17cm]{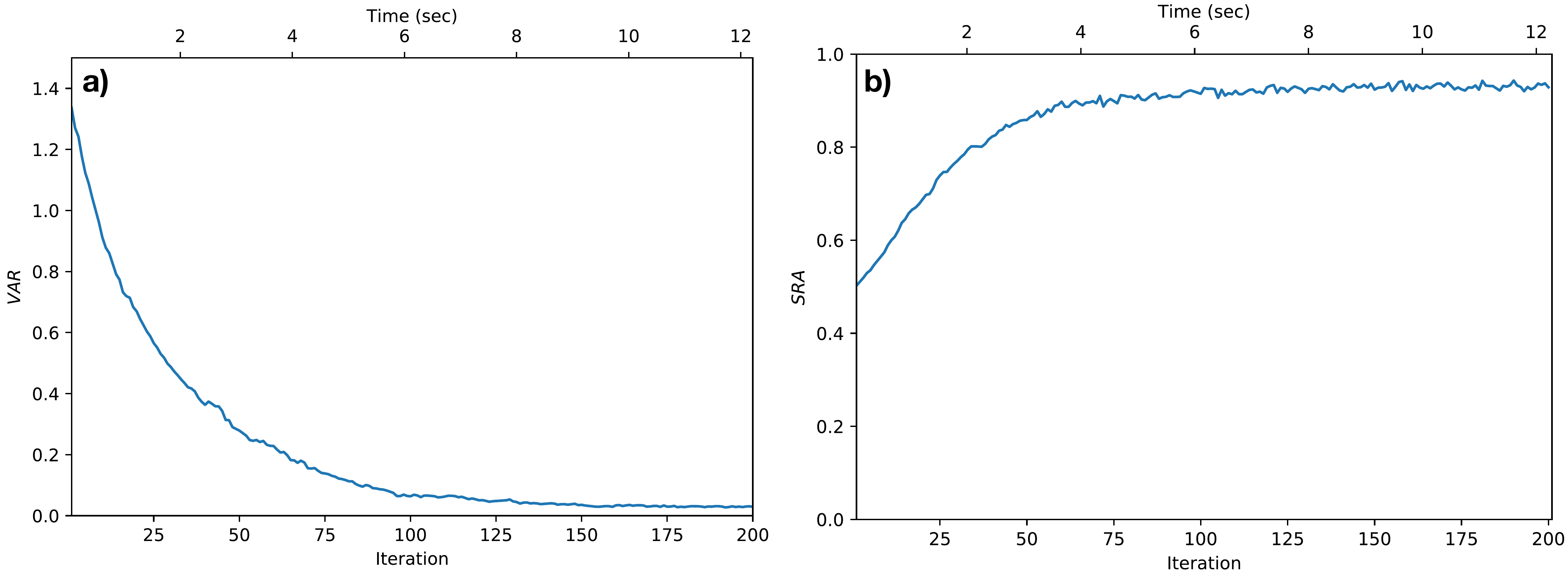}
     \caption{F\&F performance as the iterations progress. {(a) The $VAR$ as function of iteration. (b) The $SRA$ as function of iteration.}}
     \label{fig:internal_source_LWE_measurements}
\end{figure*}
\begin{figure*}
\centering
\includegraphics[width=17cm]{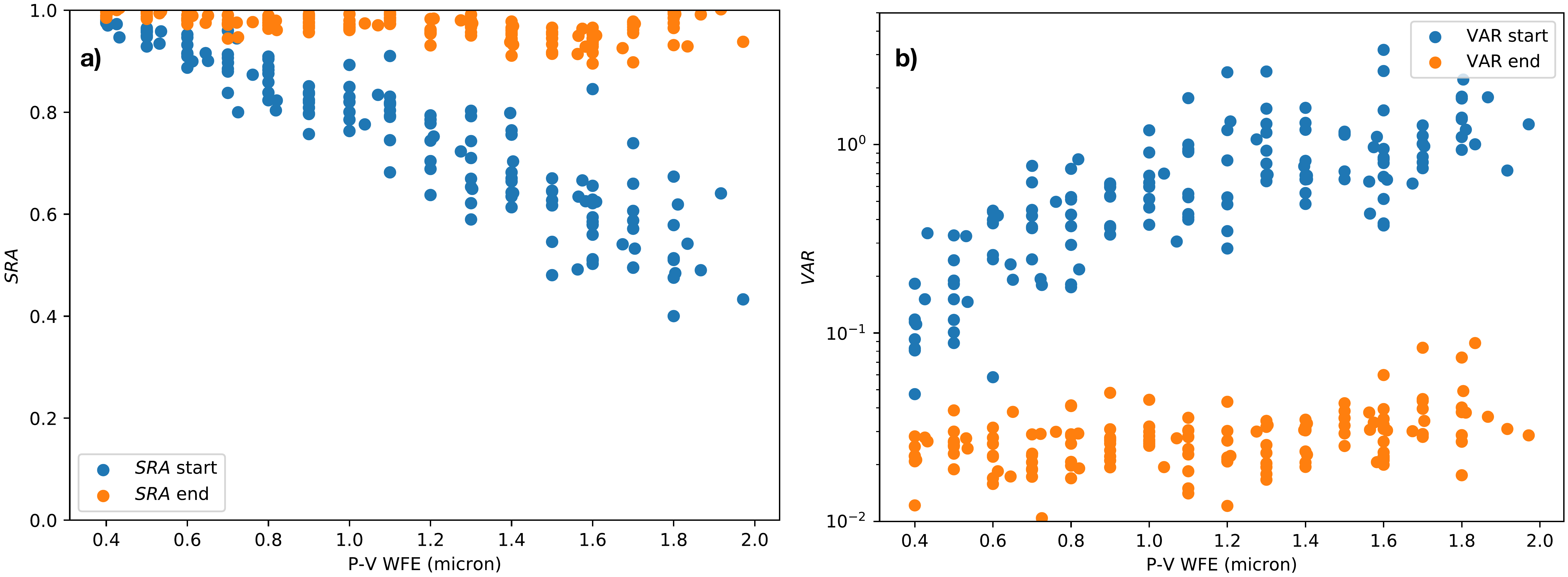}
\caption{{The $SRA$ and $VAR$ as function of the P-V WFE of the LWE} for the experiments with the internal source. Shown is the distribution before and after correction by F\&F. }
\label{fig:LWE_calibration_results}
\end{figure*}
\begin{figure}
\centering
\includegraphics[width=\hsize]{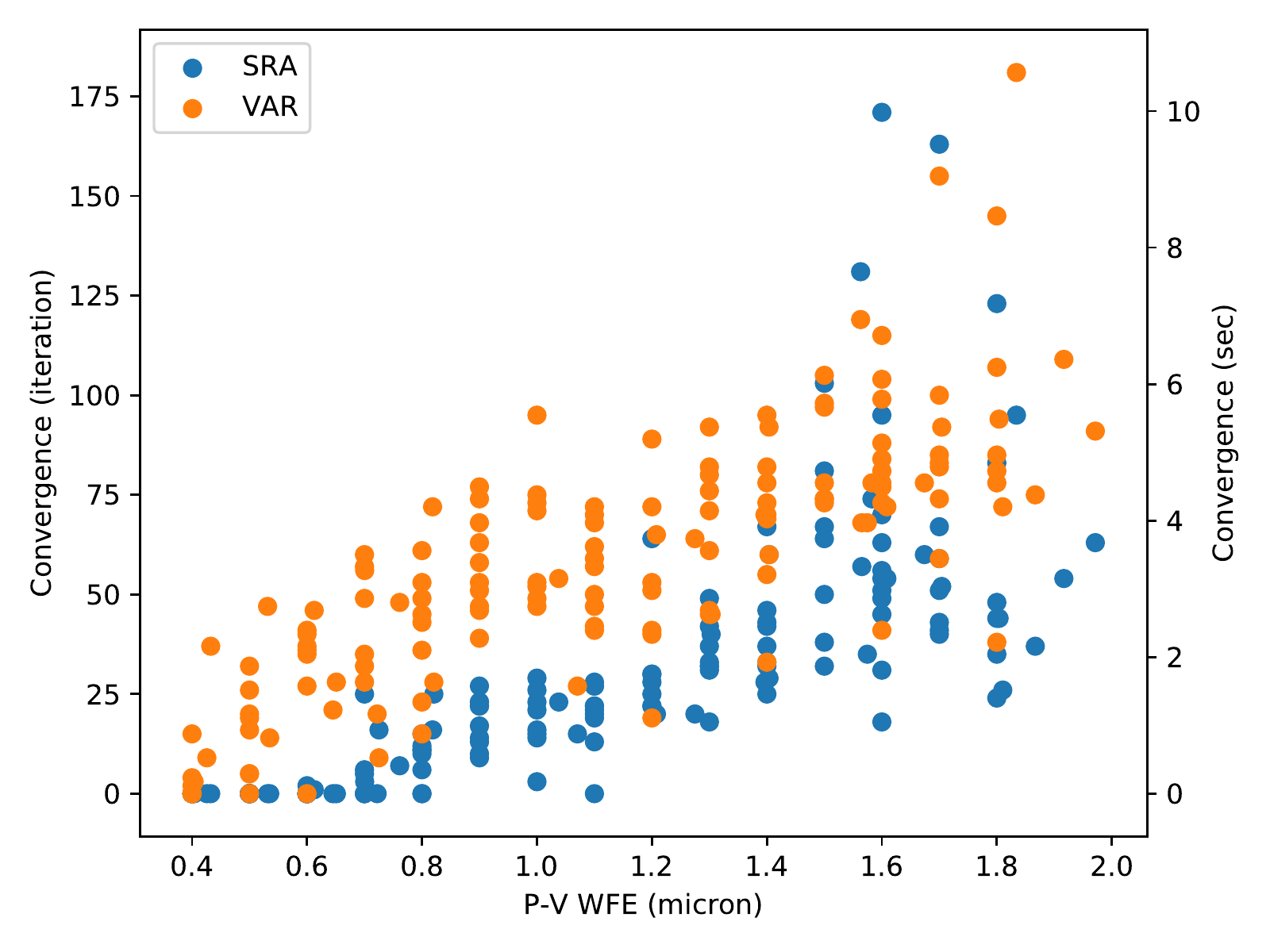}
\caption{{{Convergence time of F\&F as function of the P-V WFE of the LWE for the experiments with the internal source. The convergence time for the $SRA$ and $VAR$ was measured separately.} The algorithm converged when the $SRA > 90$ \% and the $VAR < 0.1$.}}
\label{fig:convergence}
\end{figure}
\begin{table}
\caption{Parameters of F\&F and the closed-loop settings during the internal source tests.}
\label{tab:internal_source_settings}
\vspace{2.5mm}
\centering
\begin{tabular}{l|l}
\hline
\hline
Parameter & Value \\ \hline
$\epsilon$ & $10^{-3}$\\ 
Mode basis & Zernike or PTT \\
$N_{\text{img avg}}$ & 10 \\
$g$ & $0.3$\\
$c_{lf}$ & $0.999$ \\
$N_{\text{iter}}$ & 200 \\
\hline
\end{tabular}
\end{table}
\begin{figure*}
\centering
   \includegraphics[width=17cm]{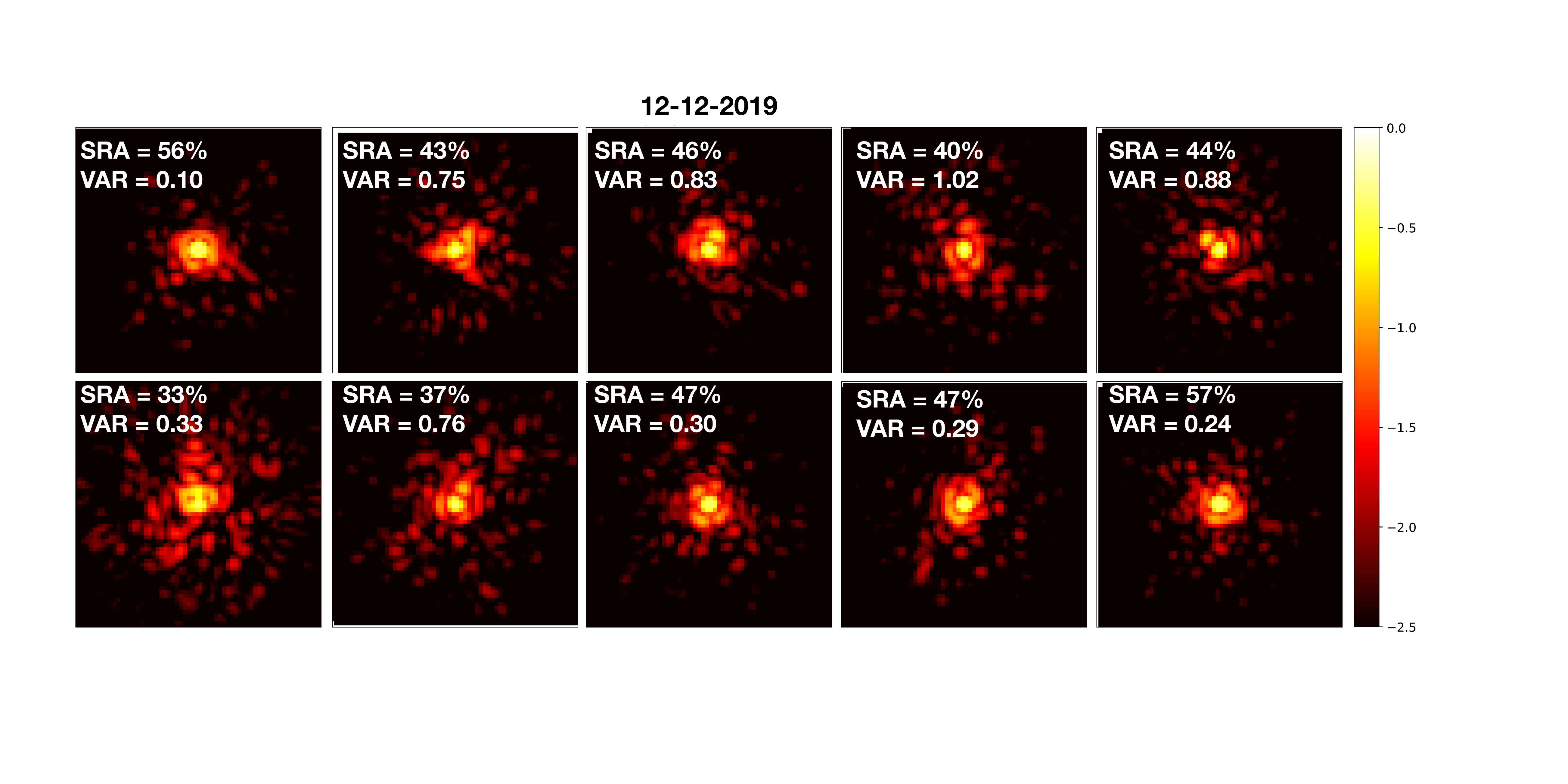}
     \caption{Short exposure images recorded during an open-loop test of F\&F. {Individual images consist of 10 aligned and stacked images, each with an integration time of 19 $\mu$s, with a total integration of 0.19 ms. These show }that the PSFs are severely distorted by the challenging atmospheric conditions. The first Airy ring is always broken up, and higher order diffraction structure is not visible. All PSFs are normalized to their maximum value, {and are plotted in logarithmic scale}.}
     \label{fig:open_loop_images}
\end{figure*}
We conducted tests with the internal source in SCExAO. 
The goal was to show that F\&F in closed-loop control can be used to measure and correct NCPA and the LWE. 
The parameters for F\&F and the closed-loop settings that were used during these tests are shown in \autoref{tab:internal_source_settings}.
{There were no other AO loops running during these tests.}
The first test was to calibrate the static aberrations in the optical path of the NIR camera. 
We used the narrow band filter ($\Delta \lambda = $ 25 nm) at 1550 nm.
As we expected optical misalignments to dominate the NCPA, we decided to project the F\&F output on the lowest 50 Zernike modes. 
In \autoref{fig:internal_source_images} a and b, the pre- and post-NCPA calibration PSFs are shown. 
The $SRA$ has increased from $94\%$ to $97\%$, the first Airy ring becomes less distorted, which is reflected in the $VAR$ going down from 0.15 to 0.03.
This shows that F\&F is suitable to correct low-order NCPA. \\
\indent The next test was to introduce a severe LWE wavefront ($1.6$ $\mu$m P-V) and correct it with the algorithm. 
Here, we chose to project the estimated phase on the PTT mode basis, as the NCPA were already compensated by the previous test, and the PTT modes were assumed to dominate. 
In \autoref{fig:internal_source_images} c and d, we show the PSF with the LWE and after the correction.
When the LWE is introduced, the PSF is heavily distorted and broken up into multiple parts. 
This is quantified by the $SRA$ of $50\%$ and the $VAR$ of $1.34$. 
After correction, the PSF is almost restored the original aberration-free version of itself, it looks very similar to \autoref{fig:internal_source_images} b with some slight vertical elongation due to an uncorrected aberration. 
Quantitatively, the $SRA$ increased to $94\%$ and the $VAR$ decreased to $0.03$. 
The evolution of the $VAR$ and $SRA$ during the test are shown, respectively, in \autoref{fig:internal_source_LWE_measurements} a and b.  
The $SRA$ and $VAR$ have mostly converged in $\sim$ $100$ iterations and {remained} stable at that level. \\

\indent We expanded the LWE correction test by including a set of LWE phase screens in a range P-V WFEs to verify that F\&F can bring back the PSF quality. 
We tested 153, random, LWE phase screens with a P-V WFE between 0.4 and 2 $\mu$m.  
For each of these phase screens, we calculated the $SRA$ and $VAR$ before and after correction, the results of which are shown in \autoref{fig:LWE_calibration_results}.
These show that, for the initial, uncorrected images, the $SRA$ decreases for increasing WFE.
The $VAR$ increases with increasing WFE, but its values have a bigger spread than the $SRA$.
For example, when the WFE is 0.9 $\mu$m P-V, the $SRA$ varies between 75\% and 90\%, while the $VAR$ fluctuates between $0.3$ and $0.6$.
Also for higher WFE, for example, at 1.8 $\mu$m P-V, the $VAR$ is distributed between 1 and 2. 
Although the $VAR$ generally increases with P-V WFE, due to the large spread, the $VAR$ on its own does not seem to be a good indicator for the amount of WFE other than that there is WFE present. 
After correction, the distributions of the $SRA$ and $VAR$ flatten to above 90\% and under $\sim$0.05, respectively.
Thus, the LWE phase screens were successfully corrected in all the tested cases.
{We also measured the convergence time of F\&F for each of the LWE phase screens.}
{The convergence time was measured separately for the $SRA$ and the $VAR$.}
{The algorithm was said to have converged when the $SRA > 90$\% and the $VAR < 0.1$.}
{In \autoref{fig:convergence}, the results are shown.}
{The convergence time goes up with increasing P-V WFE, with the $VAR$ having slightly longer convergence times.}
{For most P-V WFEs, the $SRA$ converged within 75 iterations, which corresponds to $\sim$4.5 seconds.}
{For the $VAR,$ most tests converged within 100 iterations, which is $\sim$6 seconds.}
{For some phase screens the convergence time is zero, which is because the phase screens were not severe enough to push the PSF out of the converged regime.} 

\subsection{On-sky demonstration}\label{sec:onsky} 
\begin{table}
\caption{Parameters of F\&F and the closed-loop settings during the on-sky tests.}
\label{tab:on-sky_settings}
\vspace{2.5mm}
\centering
\begin{tabular}{l|ll}
\hline
\hline
Parameter & Value (12-12-2019) & Value (30-01-2020) \\ \hline
$\epsilon$ & $10^{-2}$ & $10^{-3}$\\ 
Mode basis & Zernike + PTT & Zernike + PTT \\
$N_{\text{img avg}}$ & 10 & 10 \\
$g$ & $0.3$ & $0.3$\\
$c_{lf}$ & $0.999$ & $0.999$ \\
$N_{\text{iter}}$ & 1000  & 500 / 1000 \\
\hline
\end{tabular}
\end{table}
\begin{figure*}
\centering
   \includegraphics[width=17cm]{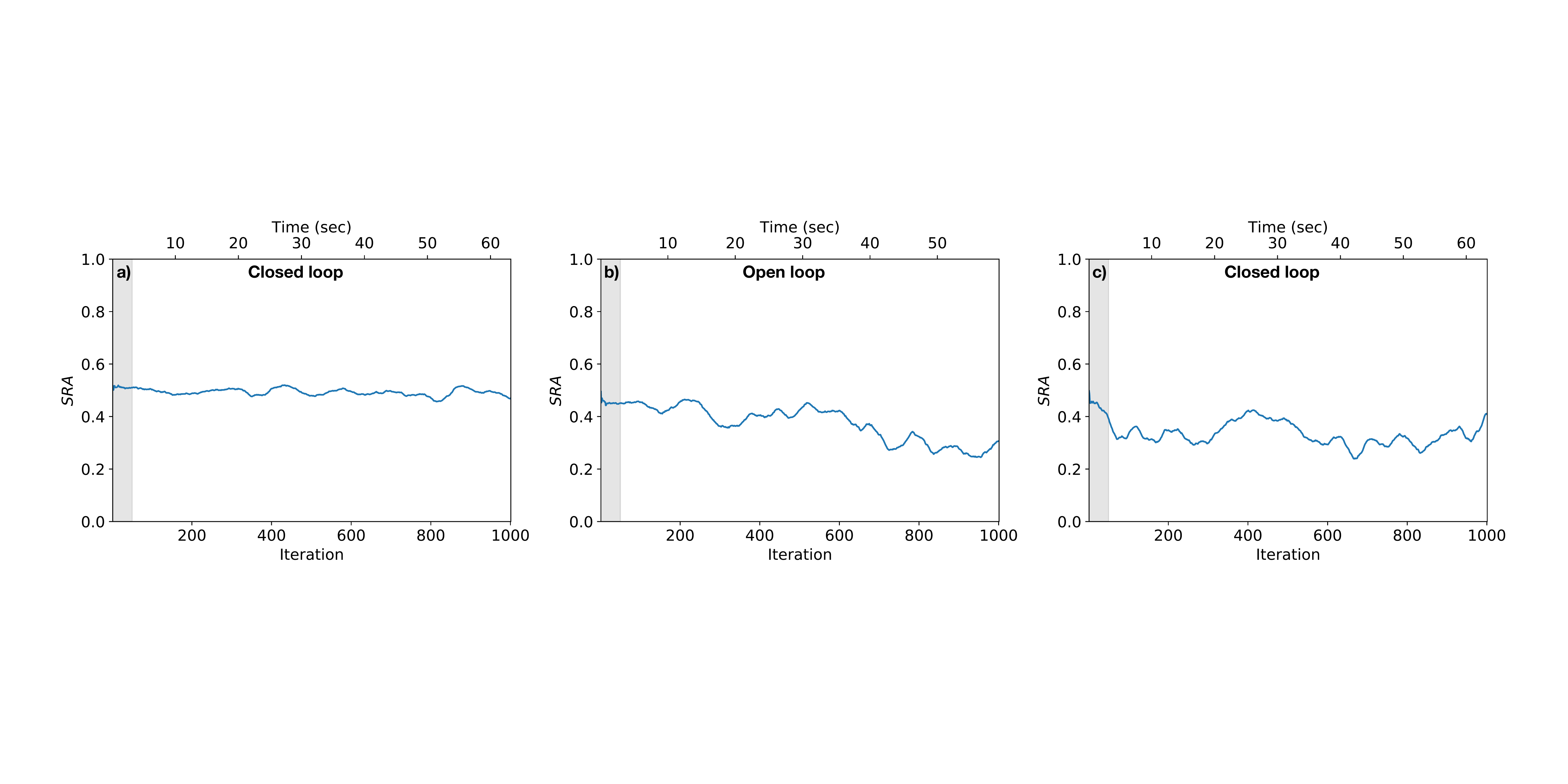}
     \caption{{Measurements of $SRA$}  on running average images during three, subsequent in time, on-sky experiments. The running average image for iteration $i$ is defined as the average of images $i-50$ to $i$. The gray box denotes the iterations for which the full average of 50 images could not be calculated. (a) The measurements during the first closed-loop test. (b) The F\&F loop was opened, meaning {the} gain was set to zero and its DM correction removed. (c) Loop was closed again.}
     \label{fig:on_sky_strehl}
\end{figure*}
\begin{figure*}
\centering
   \includegraphics[width=17cm]{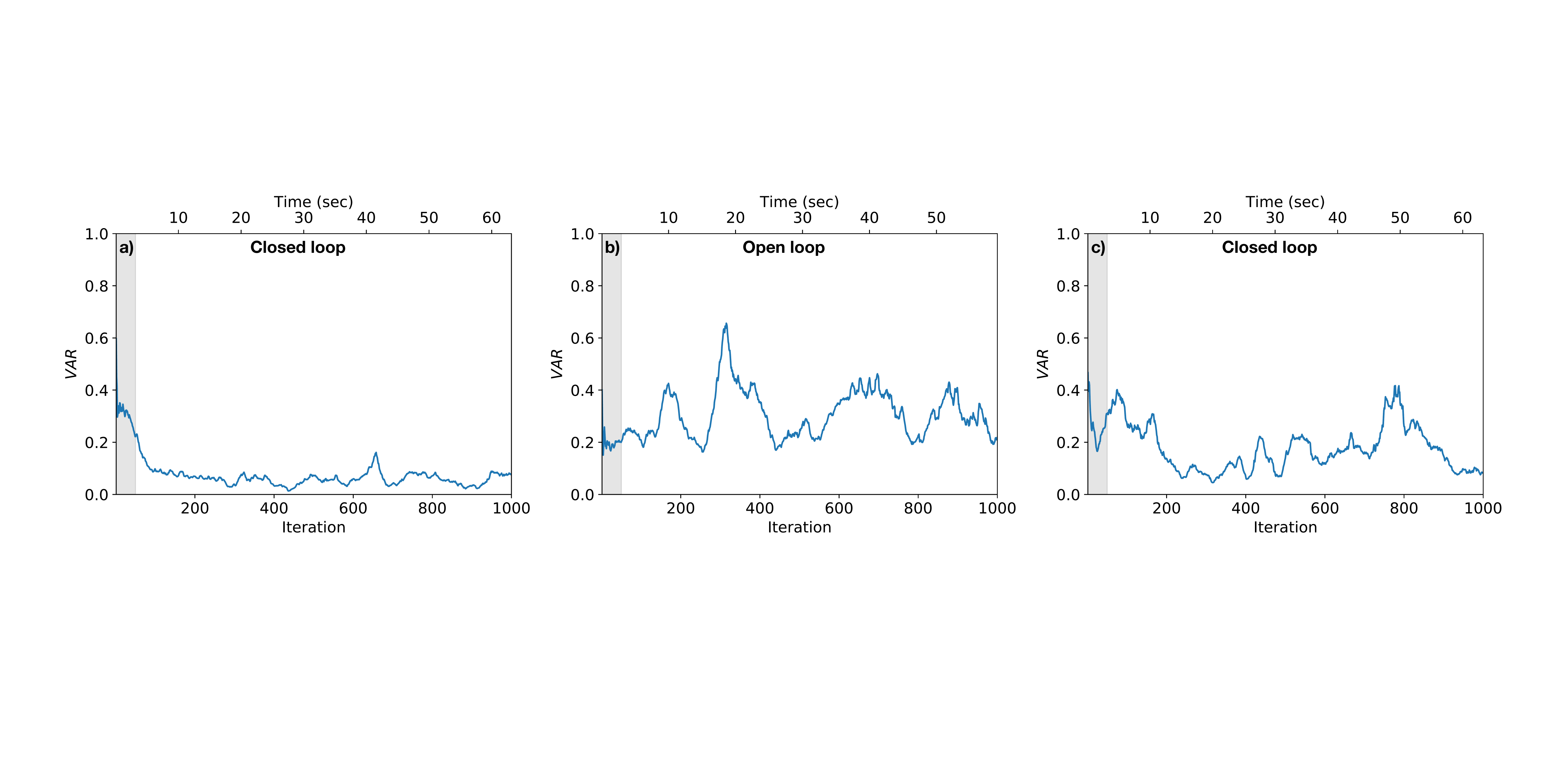}
     \caption{{Measurements of $VAR$} on running average images during three, subsequent in time, on-sky experiments. The running average image for iteration $i$ is defined as the average of images $i-50$ to $i$. The gray box denotes the iterations for which the full average of 50 images could not be calculated. (a) The measurements during the first closed-loop test. (b) The F\&F loop was opened, i.e. gain was set to zero and its DM correction removed. (c) Loop was closed again.}
     \label{fig:on_sky_asymmetry_factor}
\end{figure*}
We tested F\&F on-sky during two SCExAO engineering nights. 
The first tests were done in the first half night of December 12, 2019, while observing the bright star Mirach ($m_H = -1.65$).
The tests started at 19:24 and ended at approximately 20:00 (HST).  
The atmospheric conditions were not ideal, seeing measurements during the F\&F test {were recorded} to be between 1-1.1" in H-band{, corresponding to 1.3-1.4" seeing at 500 nm.} 
In less severe conditions, when SCExAO can deliver a good AO performance, it routinely achieves estimated Strehl ratios above 90\%\footnote{\url{https://www.naoj.org/Projects/SCEXAO/scexaoWEB/020instrument.web/010wfsc.web/indexm.html}}.
In comparison, during these tests we report a $SRA$ between 34\% and 49\%.
The individual images (that F\&F {used} for its phase estimates) were heavily distorted, for instance, the first Airy ring was always broken up, and higher order diffraction structure was not visible.
As an example, \autoref{fig:open_loop_images} shows images that were taken during open-loop measurements, {without F\&F running but with the PYWFS loop closed.} 
The wind speed of the jet stream was {forecasted} to be 22.2 m/s at 20:00 (HST)\footnote{\url{https://earth.nullschool.net/}}.
The nearby CFHT telescope (located 750 m to the east of the Subaru Telescope) reported a wind speed between 4.5 and 7 m/s during the tests\footnote{\url{http://mkwc.ifa.hawaii.edu/archive/wx/cfht/}}. 
Simultaneously, the wind speed inside the dome of the Subaru Telescope was measured to be between 0 and 0.3 m/s.
{A further analysis of all wind speed data measured in 2019 by CFHT and within the Subaru dome revealed that these were typical conditions, and therefore cannot be considered individually to indicate LWE occurrence.}
In \autoref{tab:on-sky_settings}, the settings for F\&F and the loop are shown. 
The F\&F loop was running at 12 {FPS}. 
These experiments were performed with the H-band filter, as it was already in place when the experiments started. 
It was not possible to separate NCPA and LWE calibrations, and therefore we projected the F\&F phase estimate on the combined Zernike and PTT mode basis to be able to simultaneously sense and correct them. \\

\indent Here, we present the tests where we first closed the F\&F loop, then opened it (by setting the gain to zero) and removed the DM command, and then closed the loop again.
Each of these tests was conducted with 1000 iterations. 
As shown in \autoref{fig:open_loop_images}, the individual images were severely distorted by the atmosphere. 
To suppress atmospheric effects and more accurately measure the performance of F\&F  on long exposure images, we introduced running average images. 
The running average image on iteration $i$ is defined as the average of the images $i-50$ to $i$.
The $SRA$ estimated during these tests is shown in \autoref{fig:on_sky_strehl}. 
This figure shows that during the first closed-loop tests, the $SRA$ was relatively stable around 50\%, and when the F\&F loop opened, it slowly deteriorated to below 40\%.
When the F\&F loop closed again, the $SRA$ varied between 30\% and 50\%. 
Roughly half way through the open loop and through the last closed-loop test, the atmospheric conditions started deteriorating, explaining the strong variations and loss in $SRA$.  
In \autoref{fig:on_sky_asymmetry_factor}, we show similar plots but for the $VAR$.  
These figures show that the $VAR$ was significantly lower during the closed-loop tests than during the open-loop test.
In the first closed-loop test, the $VAR$ decreased within the first hundred iterations and then remained relatively stable around $0.1$.
When the loop opened, the $VAR$ never got under $0.2,$ and it even peaked at $\sim$0.65 around three hundred iterations.
When the loop was closed again, the $VAR$ again decreased in $\sim$two hundred iterations. 
It did not remain as stable as in the first experiment, which is likely due to the deteriorated atmospheric conditions, but it is still lower than the open-loop experiment.  
The oscillations in $VAR$ observed in all three tests could be due to changes in the LWE.
\begin{figure*}
\centering
   \includegraphics[width=17cm]{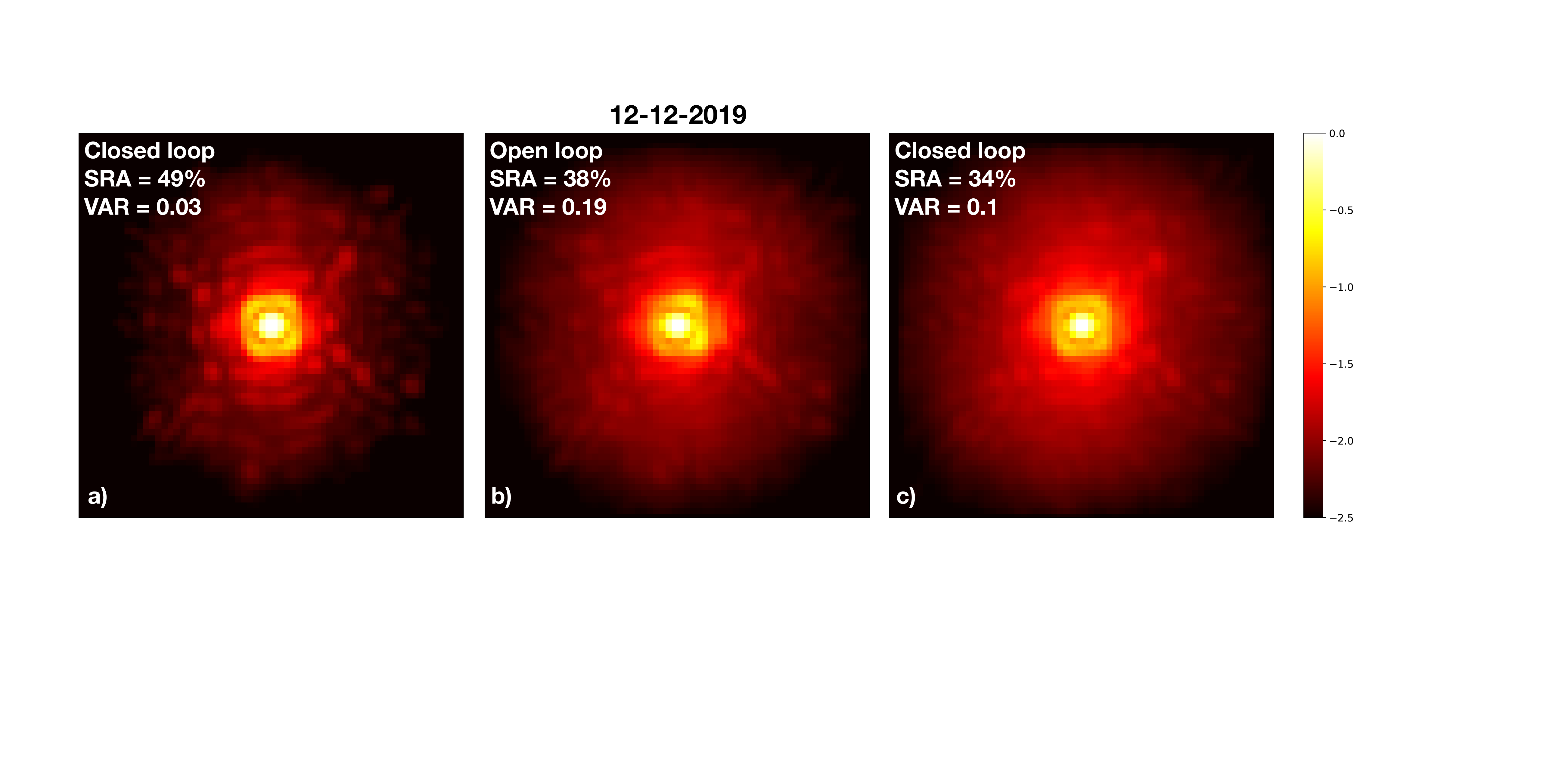}
     \caption{Averaged PSFs during during three, subsequent in time, on-sky experiments. All PSFs are normalized to their maximum value, {and are plotted in logarithmic scale}. During these experiments, the atmospheric conditions degraded, explaining the lower {$SRA$}. (a) The average PSF with a closed F\&F loop. (b) The average PSF when the F\&F loop was opened and its DM correction removed. (c) The average PSF when the F\&F loop was closed again. }
     \label{fig:on_sky_average_PSF}
\end{figure*}
Finally, in \autoref{fig:on_sky_average_PSF}, we show the PSFs that are averaged over all the iterations, and therefore suppress most of the atmospheric effects. 
These PSFs also clearly show how the deteriorating conditions, such as the halo around the PSF, which is caused by residual wavefront errors, become significantly more visible during the experiments. 
It shows that when the loop is closed, the $VAR$ converges to 0.03 - 0.10, and when the loop is open, the $VAR$ is 0.19.
This clearly shows that, even when the atmospheric conditions are challenging, F\&F manages to increase the symmetry of the PSF and thus corrects aberrations distorting the PSF.  
{This was also observed in all other tests performed during this night, which are not presented in this work.}
However, although the circumstances seemed to be right (low ground wind speed), we cannot be sure that during these tests we corrected LWE aberrations, {static aberrations upstream of SCExAO, or NCPA.} \\ 
\begin{figure*}
\centering
   \includegraphics[width=17cm]{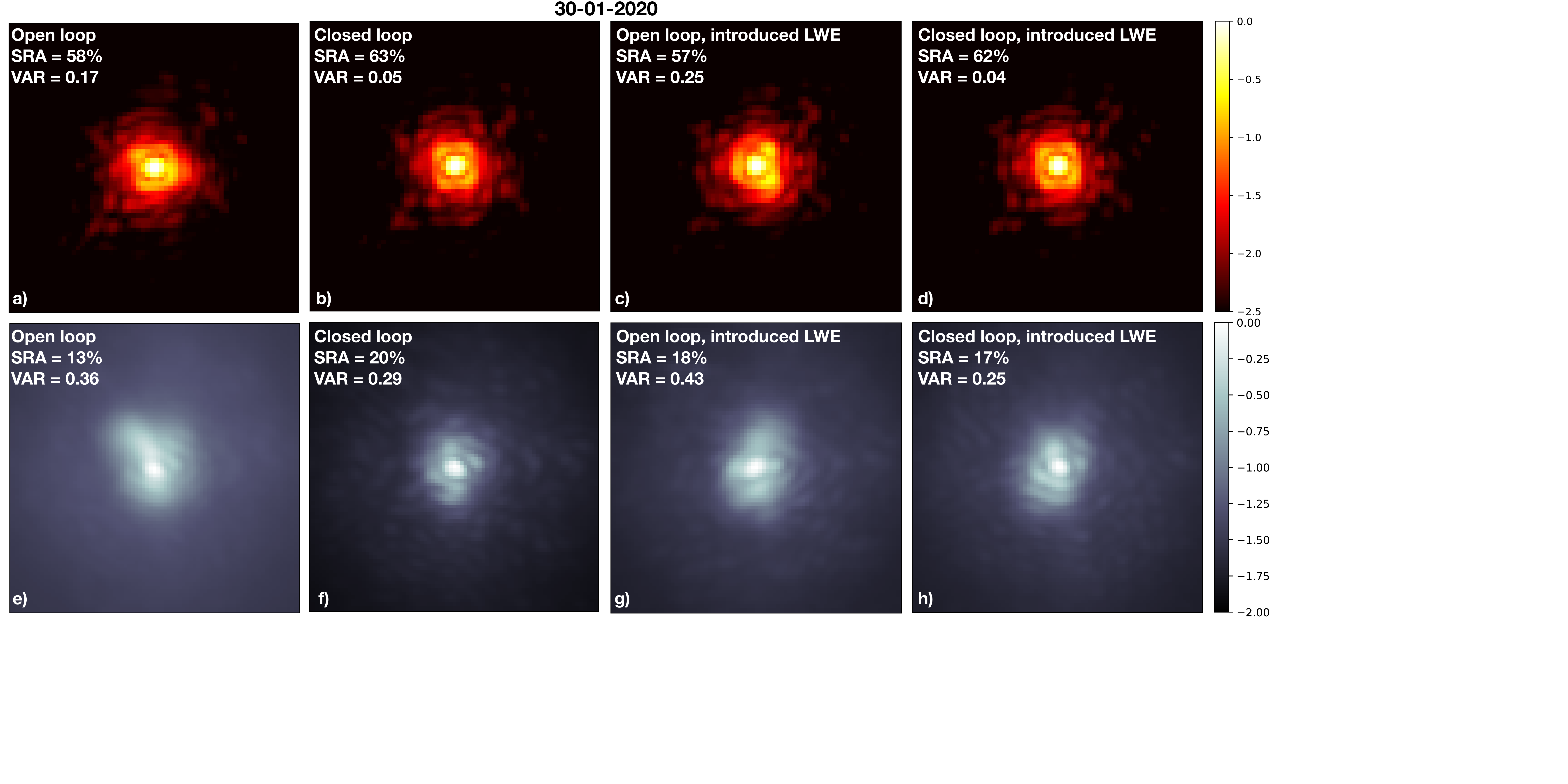}
     \caption{Averaged PSFs from four different on-sky experiments. The top row shows the PSFs in the NIR, while the bottom row shows the PSFs in the {optical}. All PSFs are normalized to their maximum value, {and are plotted in logarithmic scale}. (a) and (e): PSFs while the F\&F loop was open and no (previous) F\&F DM correction applied. {The optical PSF is significantly distorted}. (b) and (f): The PSFs while the F\&F was closed. Both PSFs improved, a clear sign that {aberrations common to both the optical and NIR path were} (partially) corrected. (c) and (g): The PSFs while the F\&F loop was opened and a LWE-like wavefront was applied by the DM, but no F\&F correction was applied. (d) and (h): Closed loop PSFs with the LWE introduced on the DM, which was successfully corrected.}
     \label{fig:on_sky_average_PSF_2}
\end{figure*}
\indent We conducted more F\&F on-sky tests during the first half night of January 30, 2020. 
We observed Rigel ($m_H=0.2$), and the tests approximately started and ended at 23:36 and 23:48 (HST), respectively.
We did not make seeing measurements, but the conditions appeared to be somewhat better than for the previous on-sky tests.  
The wind speed in the dome of the Subaru Telescope was again reported to be very low, between 0 and 0.2 m/s. 
The CFHT telescope reported a windspeed between 3 and 4 m/s. 
Again, {typical wind speed conditions.} 
The jet stream wind speed was {predicted to be between} 11 and 22 m/s, significantly higher than the ground windspeed. 
The settings of the algorithm are shown in \autoref{tab:on-sky_settings}.
The F\&F loop was running at 12 {FPS}. 
During these tests, we simultaneously recorded data in the {optical} with the VAMPIRES instrument.  
{The goal was, given the system layout \autoref{fig:system_layout}, to rule out NCPA as the corrected aberration, as a PSF improvement both in the optical and NIR would point towards corrected aberrations in the common optics.}  
{These aberrations could be (quasi-)static aberrations in the telescope and AO188, and/or the LWE.}
The VAMPIRES instrument was recording short exposure data at 200 {FPS} at 750 nm ($\Delta \lambda$ = 50 nm), and its images were aligned and stacked to get an estimate of its long exposure PSF. 
The VAMPIRES images {were} {also be analyzed using the SRA (\autoref{eq:strehl_measurement}) and VAR (\autoref{eq:VAR})}. 
{The VAR and SRA {were} calculated at $\lambda = 750$ nm, and {used} a plate scale of 6.1 mas / pixel and a clockwise rotation of 68.9$^{\circ}$ }. \\ 

\indent The two first experiments were again {with an open and closed} F\&F loop to quantify how F\&F improves the nominal PSFs. 
These tests were done for 1000 iterations of the F\&F loop and the results are shown in \autoref{fig:on_sky_average_PSF_2}. 
The NIR and {optical} PSFs are shown in \autoref{fig:on_sky_average_PSF_2} a and e, respectively.
{The NIR PSF shows an asymmetric first Airy ring, and has an $SRA$ of 58\% and a $VAR$ of 0.17.}
{The optical PSF {was} heavily distorted, almost no diffraction structure {was} observed and {was} very elongated, corresponding to an $SRA$ of 13\% and a $VAR$ of 0.36.}
When the F\&F loop {closed} (\autoref{fig:on_sky_average_PSF_2} b and f), the $SRA$ of the NIR PSF {rose} to 63\%, and the $VAR$ {dropped} to 0.05. 
The {optical} PSF also significantly {improved}: {the $SRA$ {became} 20\%, the $VAR$ {dropped} to 0.29,} the strong elongation {disappeared} and diffraction structure {became} more visible. 
{Both PSFs have improved, which is a strong sign that aberrations in the common optics got corrected, either the LWE or statics in the telescope and AO188.} 
During the next tests, we introduced a LWE-like wavefront on the DM (0.8 $\mu$m P-V) after removing the previous F\&F corrections, and recorded the open and closed-loop data. 
{The main AO loop remained closed while recording this data, and the PYWFS reference was updated in such a way that the PYWFS would not correct the LWE-like wavefront (a similar offset that is used for the F\&F loop).}
This PYWFS reference offset {was} calculated such that the DM command by the PYWFS {was} on average zero, meaning the PYWFS {was} only correcting wavefront errors from the free atmosphere. 
In the open-loop data (\autoref{fig:on_sky_average_PSF_2} c and g), the NIR PSF {was} more distorted than before, its $SRA$ {was} 56\%, and the $VAR$ {was} 0.25.
The first Airy ring {was} broken up into three bright lobes, a typical signature of the LWE. 
The {optical} PSF {was} still heavily distorted, but its elongation rotated, {and {had} a $SRA$ of 18\% and a $VAR$ of 0.43.} 
When the F\&F loop {closed} (\autoref{fig:on_sky_average_PSF_2} d and h), {it restored} the NIR PSF back to a $SRA$ of 62\% and a $VAR$ of 0.04. 
The {optical PSF also became more symmetric, as the $VAR$ decreased to 0.25, the $SRA$ stayed approximately the same at 17\%.} 
\section{Discussion and conclusion}\label{sec:conclusions} 
The Fast and Furious sequential phase diversity algorithm has been deployed to the SCExAO instrument at the Subaru Telescope.
This is in the context of measuring and correcting non-common path aberrations (NCPA), the island effect (IE), and the low-wind effect (LWE).
Both of these effects are considered to be limiting factors in the detection of exoplanets in high-contrast imaging observations. 
In this paper, we present the results of experiments both with the internal source and on-sky. 
We measured the quality of the PSF using two metrics: 1) the Strehl ratio approximation ($SRA$; \autoref{eq:strehl_measurement}), and 2) the variance of the normalized first Airy ring ($VAR$; \autoref{eq:VAR}), which measures the distortion of the first Airy ring.
Using the internal source, we tested random LWE aberrations between 0.4 and 2.0 $\mu$m and show that F\&F is able to correct these aberrations and bring the $SRA$ above 90\% and the $VAR$ below 0.05. 
Although we only managed modest improvements in PSF quality, we demonstrated during multiple on-sky tests significant gains in PSF stability.
During these tests, the F\&F loop was running at 12 {FPS}.
In the first tests, no improvement in $SRA$ was observed, which we attribute to the challenging atmospheric circumstances during these tests (seeing was {1.3-1.4'' at 500 nm}).
The $VAR,$ however, did improve from 0.19 to 0.03, indicating greater PSF stability within the control region of F\&F.  
During further on-sky tests, we did observe an $SRA$ improvement of $\sim$5\% in the NIR, but it is unclear if it can be attributed to a correction of the LWE {{and/or} static aberrations} or to changing atmospheric conditions.
The $VAR$ improved from 0.17 to 0.05 during these tests. 
Simultaneously, we also recorded the PSF in the {optical} with the VAMPIRES instrument. 
{The goal was to investigate if we were correcting aberrations common to both the optical and NIR path, or NCPA.}
When the F\&F loop was closed, the {optical PSF also significantly improved, meaning the $SRA$ increased by $\sim$7\% and the $VAR$ improved from 0.36 to 0.29.}
{These results strongly imply that we were correcting aberrations common to both paths, which could be the LWE and/or statics upstream of SCExAO.}
{Although the windspeed in the dome of Subaru was low (between 0 and 0.2 m/s), we can not conclude that we actually corrected the LWE as there were no independent measurements available.}
These tests show that F\&F is able to improve the wavefront, even during very challenging atmospheric conditions. \\
\indent {The characteristic timescale of the LWE was determined {to be} $\sim$1 to 2 seconds (\citealt{sauvage2016tackling}; \citealt{milli2018low}) in context of VLT/SPHERE.}  
{It is unclear if these timescales also apply to  Subaru/SCExAO as it has a different spider geometry.}
{If we assume that the timescales are similar, then the convergence times of F\&F presented in \autoref{fig:convergence} are not sufficient.}
{However, we foresee some improvements to the implementation of F\&F at SCExAO that would bring the convergence timescale in the regime that would allow effective LWE correction.}
{These improvements are as follows:}
\begin{enumerate}
\item {In the work presented by \cite{wilby2018laboratory}, the algorithm converged in fewer iterations ($\sim$10 iterations) than the internal source results presented in this work ($\sim$100 iterations).}
{In simulation work performed in context of SCExAO, we also found similar convergence times ($\sim$10 iterations; \citealt{vievard2019overview}).}
{This means that there is an unaccounted for gain factor in the current implementation at SCExAO.}
{If this gain factor is resolved, the the convergence time would increase by a factor of $\sim$10.}
\item {As discussed in \autoref{sec:implementation}, the current loop speed is limited by the implementation in Python, and not by the frame-rate of the NIR camera.}
{This was also the case for the on-sky tests.}
{We expect that, when the algorithm is implemented in C, 300-400 FPS would be relatively easily {achievable}.}
\item {As also discussed in \autoref{sec:implementation}, the current bottleneck in the Python implementation is the image alignment.}
{During the on-sky tests, we aligned and averaged 10 images for every iteration of F\&F.}
{If this is reduced to one image for every F\&F iteration, the loop speed would also increase by a factor of a few.}
\item For both the internal source and the on-sky tests, the loop settings and F\&F parameters (loop gain, leakage factor, and $\epsilon$) were not optimized.
For example, during the on-sky experiments $\epsilon$ (regularization parameter for odd phase modes, see \autoref{eq:odd_electric_field}) was varied between $10^{-2}$ and $10^{-3}$.
This changes the algorithm sensitivity to odd modes, but it is unclear how much it affects the on-sky performance.  
Therefore, we expect tweaking these parameters to lead to a performance gain in terms of convergence speed. 
\end{enumerate}
{These improvements will be tested in future work.} \\
\indent  {The} experiments with the internal source were carried out with the narrowband filter at 1550 nm ($\Delta\lambda = 25$ nm).
This bandwidth is relatively close to monochromatic, and thus close to the ideal performance of the algorithm as it assumes monochromatic light. 
However, the on-sky experiments were carried out using roughly half of the bandwidth of H-band, and still show satisfactory results.
Therefore, quantifying the performance difference between narrowband and broadband filters would also be of interest. \\ 
\indent The implementation of F\&F presented in this paper assumes, and therefore only estimates, phase aberrations. 
Although phase aberrations are currently limiting observations, amplitude aberrations due to the atmosphere and instrumental errors will start to limit raw contrast at the $\sim$ $10^{-5}$ level \citep{guyon2018extreme}.
Therefore, implementing the extended version of F\&F presented by \cite{korkiakoski2014fast}, which can measure both phase and amplitude will also be of interest.
We only demonstrated low-order corrections by projecting the F\&F phase estimate on the first 50 Zernike modes and the piston-tip-tilt modes, because we focused on correcting the IE. 
Higher order corrections with F\&F are possible \citep{korkiakoski2014fast}, but will need to be tested on sky. \\
\indent For F\&F to be operated effectively and routinely during high-contrast imaging observations, the algorithm needs to be integrated in the system in such a way that it can run simultaneously with the coronagraphic mode.  The algorithm would preferably have access to a focal plane as close as possible to the science focal plane, as it will also correct the NCPA as much as possible.  
The most important limitation is that F\&F needs a pupil-plane electric field that is (close to) to real and symmetric, and that there is no focal-plane mask. 
The coronagraph with which the algorithm can most easily be integrated is the shaped pupil coronagraph \citep{kasdin2007shaped}. 
This coronagraph suppresses starlight by modifying the pupil-plane electric field with symmetric amplitude masks. 
Therefore, F\&F is expected to be able to operate on the PSF generated by a shaped pupil coronagraph. 
Another coronagraph in which F\&F can be integrated is the vector-Apodizing Phase Plate (vAPP; \citealt{snik2012vector}; \citealt{otten2017sky}). 
The vAPP has been deployed to multiple instruments (MagAO; \citealt{otten2017sky}, MagAO-X; \citealt{miller2019spatial}, SCExAO; \citealt{doelman2017patterned}, LBT; \citealt{doelman2017patterned}, and LEXI; \citealt{haffert2018sky}). 
The vAPP suppresses starlight by manipulating the pupil-plane phase and creates multiple coronagraphic PSFs. 
However, this process is never 100\% efficient, and thus there is always a non-coronagraphic PSF at a lower intensity.   
The morphology of the non-coronagraphic PSF would only depend on the shape of the pupil, and would therefore be suitable for F\&F. 
Some of these vAPPs already have other implementations of wavefront sensing (\citealt{wilby2017coronagraphic}; \citealt{bos2019focal}; \citealt{miller2019spatial}), but F\&F would be a useful addition. 
For coronagraphs that have focal-plane masks to block starlight, there are a few ways to implement F\&F (assuming that for these coronagraphs the pupil-plane electric field stays symmetric and real). 
One of these, extensively discussed in \cite{wilby2018laboratory} in the context of the SPHERE system, is to extract light for the beam just before it hits the focal-plane mask using, for example, a beam splitter. 
A way to circumvent the focal-plane mask would be to generate PSF copies of the star that are not affected by the focal-plane mask, using diffractive elements in the pupil (\citealt{sivaramakrishnan2006astrometry}; \citealt{marois2006accurate}; \citealt{jovanovic2015artificial}). 
These PSF copies can then serve as input PSFs for F\&F. \\ 
\indent In this paper, we show that F\&F is able to increase the PSF quality, both on-sky and with the internal source in SCExAO.
{Using the internal source, we show that F\&F can measure and correct a wide range of LWE- and IE-like aberrations.}
{With future algorithm upgrades and further on-sky tests, we hope to conclusively show on-sky correction of the LWE and IE.}
For future giant segmented mirror telescopes, the IE is expected to become even more significant as the support structures become wider and more numerous, and the segments have to be co-phased.
Going forward, it is suitable for incorporation into observing modes, {enabling {PSFs} of higher quality and stability} during science observations.

\begin{acknowledgements} 
{The authors thank the referee for the comments that improved the manuscript.}
The research of S.P. Bos, and F. Snik leading to these results has received funding from the European Research Council under ERC Starting Grant agreement 678194 (FALCONER).
The development of SCExAO was supported by the Japan Society for the Promotion of Science (Grant-in-Aid for Research \#23340051, \#26220704, \#23103002, \#19H00703 \& \#19H00695), the Astrobiology Center of the National Institutes of Natural Sciences, Japan, the Mt Cuba Foundation and the director's contingency fund at Subaru Telescope.
The authors wish to recognize and acknowledge the very significant cultural role and reverence that the summit of Maunakea has always had within the indigenous Hawaiian community. 
We are most fortunate to have the opportunity to conduct observations from this mountain. 
This research made use of HCIPy, an open-source object-oriented framework written in Python for performing end-to-end simulations of high-contrast imaging instruments \citep{por2018hcipy}.
This research used the following Python libraries: Scipy \citep{jones2014scipy}, Numpy \citep{walt2011numpy}, and Matplotlib \citep{Hunter:2007}.
\end{acknowledgements}

\bibliography{report} 
\bibliographystyle{aa}

\end{document}